\documentclass[iop]{emulateapj}

\shorttitle{A Nearly Naked Supermassive Black Hole}
\shortauthors{Condon et al.}

\begin{document}

\title{A Nearly Naked Supermassive Black Hole}

\author{J.~J.~Condon}
\affil{National Radio Astronomy Observatory\altaffilmark{1}, 
Charlottesville, VA 22903-2475, USA}
\email{jcondon@nrao.edu}
\author{Jeremy~Darling}
\affil{Center for Astrophysics and Space Astronomy, Department of 
Astrophysical and Planetary Sciences, University of Colorado, 389 UCB,
Boulder, CO 80309-0389, USA}
\author{Y.~Y.~Kovalev}
\affil{Astro Space Center of Lebedev Physical Institute, 
Profsoyuznaya 84/32, 117997 Moscow, Russia}
\affil{Max-Planck-Institute for Radio Astronomy, Auf dem H\"ugel 69, 
D-53121, Germany}
\and
\author{L.~Petrov}
\affil{Astrogeo Center, 7312 Sportsman Dr, Falls Church, VA 22043, USA}

\altaffiltext{1}{The National Radio Astronomy Observatory is a facility
of the National Science Foundation operated under cooperative agreement
by Associated Universities, Inc.}

\begin{abstract}
During a systematic search for supermassive black holes (SMBHs) not in
galactic nuclei, we identified the compact symmetric radio source
B3~1715+425 with an emission-line galaxy offset $\approx 8.5
\mathrm{~kpc}$ from the nucleus of the brightest cluster galaxy (BCG)
in the redshift $z = 0.1754$ cluster ZwCl~8193.  B3~1715+425 is too
bright (brightness temperature $T_\mathrm{b} \sim 3 \times 10^{10}
\mathrm{~K}$ at observing frequency $\nu = 7.6 \mathrm{~GHz}$) and too
luminous (1.4~GHz luminosity $L_\mathrm{1.4\,GHz} \sim 10^{\,25}
\mathrm{~W~Hz}^{-1}$) to be powered by anything but a SMBH, but its
host galaxy is much smaller ($\sim 0.9 \mathrm{~kpc} \times 0.6
\mathrm{~kpc}$ full width between half-maximum points) and optically
fainter (R-band absolute magnitude $M_\mathrm{r} \approx -18.2$) than
any other radio galaxy.  Its high radial velocity $v_\mathrm{r}
\approx 1860 \mathrm{~km~s}^{-1}$ relative to the BCG, continuous
ionized wake extending back to the BCG nucleus, and surrounding debris
indicate that the radio galaxy was tidally shredded passing through
the BCG core, leaving a nearly naked supermassive black hole fleeing
from the BCG with space velocity $v \gtrsim 2000 \mathrm{~km~s}^{-1}$.
The radio galaxy has mass $M \lesssim 6 \times 10^{\,9}\, M_\odot$ and
infrared luminosity $L_\mathrm{IR} \sim 3 \times 10^{11}\, L_\odot$
close to its dust Eddington limit, so it is vulnerable to further mass
loss from radiative feedback.
\end{abstract}


\section{Introduction}

The term ``active galactic nucleus'' (AGN) reflects the fact that the
supermassive black hole (SMBH) powering an AGN is normally found in
the nucleus of its host galaxy because dynamical friction against the
surrounding stars eventually drives any offset SMBH to the bottom of
the galaxy's gravitational potential well.  However, the stellar bulge
of nearly every massive galaxy contains a SMBH \citep{fer05}, so
heirarchical merging of galaxies should produce stellar bulges
initially containing two or more SMBHs that are offset in position and
velocity.  Following a merger, a pair of inspiraling SMBHs may spend
$\sim 10^{\,8} \mathrm{~yr}$ at separations $\gtrsim 1 \mathrm{~kpc}$
\citep{beg80} before forming a tight binary that could either stall at
$\sim 1 \mathrm{~pc}$ separation or merge.  Anisotropic gravitational
radiation emitted during a SMBH merger may kick the merged SMBH out of
the nucleus into a nearly radial orbit with kpc size for up to $\sim
10^{\,8} \mathrm{~yr}$ before it settles down \citep{hof07, gua08}.
Recoiling SMBHs on highly eccentric orbits spend most of their time
with velocity offsets too small to detect spectroscopically, so
\citet{gua08} suggested ``an alternative approach would be to search
for linear displacements between the AGN emission and the peak of the
stellar surface brightness.''  Finally, \citet{gov94} proposed the
existence of ``wandering'' SMBHs: ``If a galaxy is tidally disrupted
before coalescence is completed, a possibility exists that its BH will
remain naked'' for a few Gyr.  We have found what we believe is the
first example of a nearly naked SMBH in the surviving core of a
tidally stripped galaxy.

The discovery program that found the offset SMBH is outlined in
Section~\ref{sec:finding}.  The radio, optical, and infrared
identifications of the offset SMBH and its host galaxy are presented
in Section~\ref{sec:identifications}, and their long-slit optical
spectra are described in Section~\ref{sec:spectra}.
Section~\ref{sec:interpretation} gives our interpretation of the data
in terms of a nearly naked SMBH clothed only by the compact remnant of
a radio galaxy that was tidally shredded by the brightest cluster
galaxy (BCG) in the cluster \object{ZwCl 1715.5+4229} (B1950
coordinate name).  This cluster is usually called \object{ZwCl 8193}
in the literature.

\section{Finding Offset SMBHs}
\label{sec:finding}
To discover SMBHs with small linear offsets from their host
  galaxy bulges, we need a large sample of nearby galaxies containing
  SMBHs whose existence and positions are revealed by their compact
  $(\theta \lesssim 1 \mathrm{~mas})$ radio emission.

Condon, Broderick, \& Yin (in prep) compiled a sample of all Two
Micron All-Sky Survey (2MASS) Extended Source Catalog (XSC) galaxies
\citep{jar00} north of $\delta = -40^\circ$, at Galactic latitudes
$\vert b \vert > 5^\circ$, and brighter than $K_\mathrm{20fe} = 12.25$
at $\lambda = 2.16 \,\mu\mathrm{m}$, a wavelength at which luminosity
is a good tracer of total stellar mass \citep{mad98} and dust
extinction is small enough that the 2MASS positions should be close to
the bulge centroids.  The XSC rms position errors are
$0\,\farcs 3$--$0\,\farcs 5$ in each coordinate, but we found that the 2MASS
Point-Source Catalog (PSC) fitted positions better locate the bulge
centers with rms errors $\lesssim 0\farcs1$ in each
coordinate.  At the typical angular-size distance $\langle
D_\mathrm{A} \rangle \sim 200 \mathrm{~Mpc}$ of these bright galaxies,
a $\Delta \sim 1 \mathrm{~kpc}$ projected linear offset yields an
easily detectable angular offset $ \Delta \theta \sim 1''$.  Condon,
Yin, \& Broderick also
identified the NRAO VLA Sky Survey (NVSS) \citep{con98} radio sources
that have NVSS/2MASS position differences $\lesssim 15''$.

We observed the stronger NVSS identifications withq the Very Long
Baseline Array (VLBA) at X band ($\nu = 7.6 \mathrm{~GHz}$, $\lambda
\approx 4\,\mathrm{cm}$) and we re-analyzed archival VLBA data at 8.4
and 2.3~GHz in a large program covering 1215 radio sources with $S
\geq 50 \mathrm{~mJy}$ at $1.4 \mathrm{~GHz}$ to image the compact
components powered by SMBHs in or near bulge centroids (Petrov et
al.~2016, in preparation).

\begin{figure}[htb!]
\includegraphics[angle=0, scale=0.5,clip=true, 
trim = 12mm 20mm 3mm 20mm]{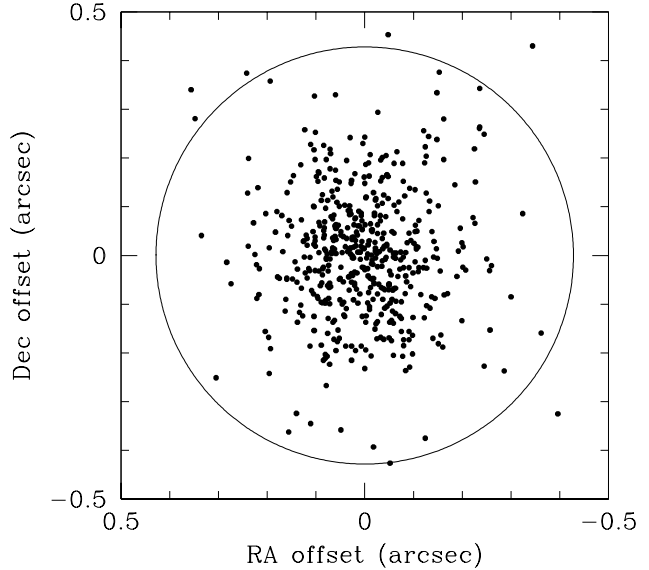}
% trim l,b,r,t
\caption{VLBA (SMBH) minus 2MASS PSC (stellar bulge) position offsets
  of 492 nearby galaxies detected by the VLBA before 2014.  The circle
  with radius $0\,\farcs 428$ should contain 99.9\% of the galaxies if
  there are no offset SMBHs and the offsets are Rayleigh distributed
  with $\sigma = 0\,\farcs115$.  Most of the points just outside the
  circle are large spiral galaxies whose stellar bulges are partially obscured
  by dust at $\lambda = 2.16 \,\mu \mathrm{m}$.  B3 1715+425 lies well
  outside the plotted region.
  \label{fig:offsets}}
\end{figure}

The 2MASS Point-Source Catalog (PSC) absolute radial position errors
are Rayleigh distributed with $\sigma \approx 0\,\farcs 115$ and our
VLBA absolute positions have $\sigma \approx 0\,\farcs001$, so SMBHs
offset by more than $ \Delta \theta \sim 0\,\farcs4$ should be
distinguished from nuclear SMBHs with 99.9\% reliability.  
The observed  VLBA minus PSC
offsets of SMBHs detected before 2014 are consistent with the expected
position uncertainties (Figure~\ref{fig:offsets}), confirming that
nearly all SMBHs are truly nuclear.  However, the compact
flat-spectrum radio source \object{B3 1715+425} (B1950 coordinate
name) is offset by $\Delta \theta > 2''$ from the infrared source
\object{2MASX 17171926+4226571} (J2000 coordinate name).

\section{Radio, Optical, and Infrared Identifications in ZwCl~8193}
\label{sec:identifications}

The BCG in ZwCl~8193 is the 2MASS galaxy 2MASX~17171926+4226571 at PSC
position J2000 $\alpha = 17^\mathrm{h}\,17^\mathrm{m}\,19\,\fs 245$,
$\delta = +42^\circ\,26'\,57\,\farcs 48$.  Its $S_{1.4\,\mathrm{GHz}}
= 133.5 \pm 4.0 \mathrm{~mJy}$ NVSS counterpart \object[NVSS
  J171719+422659]{NVSS J171719\allowbreak+422659} at J2000 $\alpha =
17^\mathrm{h}\,17^\mathrm{m}\, 19\,\fs 18 \pm 0\,\fs 04$, $\delta =
+42^\circ\,26'\,59\,\farcs6 \pm 0\,\farcs6$ is $> 2''$ north of the
BCG.  This radio source was first detected at $S = 0.2 \mathrm{~Jy}$
in the 408 MHz Bologna B3 survey \citep{fic85} and named B3 1715+425
(equinox B1950 IAU format name).  Its 4.85~GHz counterpart is \object[87GB
  171544.7+423023]{87GB 1715+4230} (B1950 name) with flux density $S =
134 \pm 16 \mathrm{~mJy}$ \citep{gre91}.  Its flat spectral index
$\alpha(1.4,\,4.85) \approx 0$ is usually a signature of synchrotron
self-absorption in a high-brightness ($T_\mathrm{b} \sim 10^{11}
\mathrm{~K}$) radio component.

\citet{all92} associated 87GB 1715+4230 with the BCG in
ZwCl~8193, for which they reported both a stellar absorption-line
redshift $z_\mathrm{a} = 0.1754 \pm 0.0006$ and a significantly higher
emission-line redshift $z_\mathrm{e} = 0.1829 \pm 0.0001$.  In a flat
$\Lambda$CDM universe with $H_0 = 70 \mathrm{~km~s}^{-1}
\mathrm{~Mpc}^{-1}$ and $\Omega_\mathrm{M} = 0.3$, $z_\mathrm{a} =
0.1754$ implies that the BCG in ZwCl~8193 is at angular-size distance
$D_\mathrm{A} \approx 613 \mathrm{~Mpc}$ and $1'' \approx 2.97
\mathrm{~kpc}$. The corresponding luminosity distance is $D_\mathrm{L}
= (1+z_\mathrm{a})^2 D_\mathrm{A} \approx 850 \mathrm{~Mpc}$.

Because ZwCl~8193 is a bright X-ray source and ``emission-line
  nebulae are frequently found surrounding the central galaxy in
  clusters with cooling flows,'' \citet{all92} believed ``that
  photoionization by an active nucleus or other activity associated
  with the radio emission is unlikely to be the principal mechanism
  for generating the line emission.'' Consequently they did not search
  for the separate radio galaxy responsible for the offset
  emission-line redshift.  \citet{wil06} presented an integral-field
  H$\alpha$+[N\,\textsc{II}] line-intensity image showing emission
  north of the BCG, but they did not discuss the radio source.
  \citet{ode10} first noted the compact radio source ``is 3
  arcsec from the center of the BCG at the location of FUV-bright
  debris features and may be associated with a merging galaxy,'' but
  associated the diffuse Ly$\alpha$ emission with ZwCl~8193. Thus the
  literature has remained unclear and the optical identification of
  B3~1715+425 is still listed as ZwCL~8193 in the NASA/IPAC
Extragalactic Database (NED).

\subsection{VLBA Images of B3~1715+425}

\citet{bou11} observed B3~1715+425 = J1717+4226 with the VLBA+European
VLBI Network at 8.4 and 2.3~GHz on 2008 March 7, and \citet{pet11}
determined its position with milliarcsecond accuracy using these
data. 

\begin{figure}[htb!]
\epsscale{1.}
\plotone{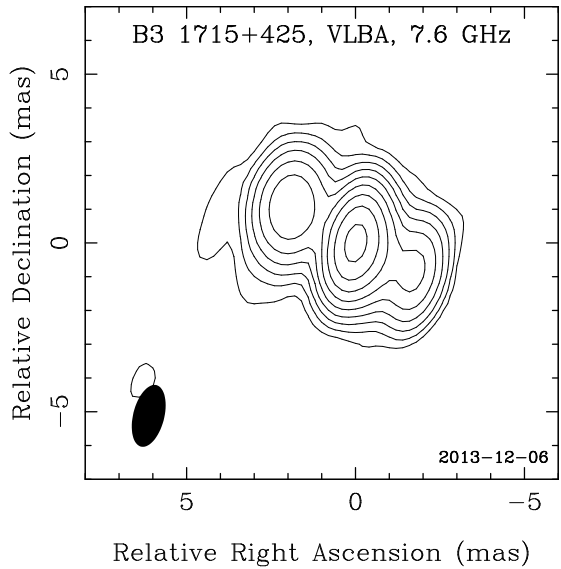}
\caption{VLBA X-band (7.6 GHz) natural-weighting image of B3
  1715+425 with contours plotted at $ 156 \, \mu\mathrm{Jy~beam}^{-1}
  \times 1$, 2, 4, 8, 16, 32, 64, 128, and 256.  The recovered flux
  density is $77 \mathrm{~mJy}$ and the core peak intensity is $52
  \mathrm{~mJy~beam}^{-1}$.  The restoring beam half-power ellipse
  ($0\, \farcs 00188 \times0\,\farcs 00092$ in $PA = -14^\circ$) is
  drawn in the lower left corner. \label{fig:vlbax}}
\end{figure}

\begin{figure}[htb!]
\epsscale{1.}
\plotone{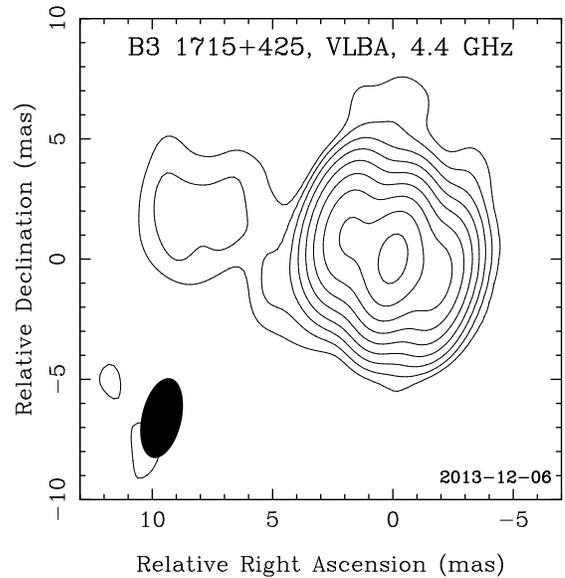}
\caption{VLBA C-band (4.4 GHz) natural-weighting image of B3
  1715+425 with contours plotted at $ 129 \, \mu\mathrm{Jy~beam}^{-1}
  \times 1$, 2, 4, 8, 16, 32, 64, 128, and 256. The recovered flux
  density is $84 \mathrm{~mJy}$ and the core peak intensity is $43
  \mathrm{~mJy~beam}^{-1}$.  The restoring beam half-power ellipse
  ($0\, \farcs 00339 \times0\,\farcs 00166$ in $PA = -13^\circ$) is
  drawn in the lower left corner. \label{fig:vlbac}}
\end{figure}

We used eight of the ten VLBA stations to reobserve B3 1715+425 in
left circular polarization (LCP) at 7.6~GHz (X band) and 4.4~GHz (C
band, with the new wideband receiver for improved sensitivity)
simultaneously for 4 hours on 2013 December 6. The missing stations
were FD (broken motor) and KP (no fringes for an unknown reason).
Half of the IFs were centered on 4.4~GHz and half on 7.6~GHz, and the
total data rate was 2048 Mbit/s.  Our $7.6 \mathrm{~GHz}$ image of B3
1715+425 is shown in Figure~\ref{fig:vlbax} and our $4.4
\mathrm{~GHz}$ image is shown in Figure~\ref{fig:vlbac}.  The position
of the strong central component in this compact symmetric radio source
derived from all VLBI observations is J2000 $\alpha =
17^\mathrm{h}\,17^\mathrm{m}\,19\,\fs 209641$, $\delta =
+42^\circ\,26'\,59\,\farcs84514$ with rms uncertainties $\sigma_\alpha
= 0\,\fs00001$ and $\sigma_\delta = 0\,\farcs0002$, a highly
significant $0\,\farcs40 \pm 0\,\farcs12$ west and $2\,\farcs37\pm
0\,\farcs12$ north of the 2MASS PSC position.

The central component has a peak intensity $S_\mathrm{p}= 52
\mathrm{~mJy~beam}^{-1}$ in our $7.6 \mathrm{~GHz}$ VLBA image.  We
approximated the central component by a circular Gaussian and directly
fit the observed fringe amplitude distribution as a function of
projected baseline length in the $(u,v)$ plane to determine its
angular diameter.  With this technique and our low rms noise levels
$\sim 0.2 \mathrm{~mJy~beam}^{-1}$ at X band and $\sim 0.1
\mathrm{~mJy~beam}^{-1}$ at C band, we could resolve the central
component by the \citet{kov05} criterion and measure its FWHM diameter
$\theta \sim 0\,\farcs 0002$.  The corresponding quantities at $4.4
\mathrm{~GHz}$ are $S_\mathrm{p} = 43 \mathrm{~mJy~beam}^{-1}$ and
$\theta \sim 0 \, \farcs 0004$.  The central component has rest-frame
luminosity $L_\mathrm{9\,GHz} \approx 6 \times 10^{\,24}
\mathrm{~W~Hz}^{-1}$ calculated from our 7.6~GHz measurements, linear
diameter $ \sim 1 \mathrm{~pc}$, and rest-frame brightness temperature
$T_\mathrm{b} \sim 3 \times 10^{10} \mathrm{~K}$.

\begin{figure}[htb!]
\begin{center}
\includegraphics[angle= -90,scale= 0.4]{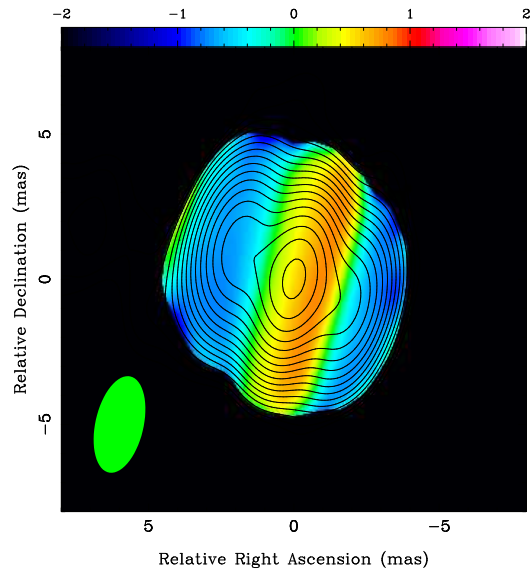}
\end{center}
\caption{Matched-resolution VLBA spectral-index image of B3
  1715+425 uses the C-band natural-weighting restoring beam 
  shown in Figure~\ref{fig:vlbac} at both
  frequencies (green ellipse). The spectral index $\alpha(4.4,\,7.6)$
  is encoded by the color bar on the top.   \label{fig:vlbaspindex}}
\end{figure}

Figure~\ref{fig:vlbaspindex} shows how the spectral index
$\alpha(4.4,\,7.6)$ varies across B3 1715+425.  To construct a
spectral-index image, we restored both the C- and X-band images with
the C-band natural-weight beam and aligned them on the northeastern
optically thin region using the two-dimensional cross-correlation
technique described by \citet{hov12} and \citet{pus12}.  At 4.4~GHz
the northeastern jet and southwestern counterjet flux densities are 21
mJy and 12 mJy, respectively; at 7.6~GHz they are 14 mJy and 8 mJy.

The spectral index of the central component was $\alpha(2.3, 8.4) =
-0.26 \pm 0.07$ at epoch 2008.2 \citep{pet11} and $\alpha(4.4, 7.6) =
+0.24 \pm 0.13$ at epoch 2013.9, indicating synchrotron
self-absorption in a high-brightness core.  The spectral-index maximum
$\alpha \sim +0.75$ appears $\sim 0\,\farcs 001$ southwest of the
nucleus.  The symmetric jets or lobes have the ``normal'' steep
spectra $-0.9 < \alpha(4.4,\,7.6) < -0.7$ of transparent synchrotron
sources.  We associate the central radio component with the apparent
base of the jet near the SMBH itself \citep{bla79} because of its
compactness, high brightness temperature, and flat spectrum.  Its
brightness temperature and spectrum are consistent with a
self-absorbed synchrotron source having a nearly equipartition
magnetic field strength, as discussed by \citet{rea94}.  The radio
source is too luminous and far too bright to be powered by anything
but a SMBH.  All known star-forming galaxies have $\nu \approx 8
\mathrm{~GHz}$ brightness temperatures $T_\mathrm{b} < 10^{\,4.5}
\mathrm{~K}$ \citep{con91}.

The symmetric source in Figure~\ref{fig:vlbax} is $\sim
0\,\farcs005 \approx 15 \mathrm{~pc}$ in projected length, so
B3~1715+425 may be a compact symmetric object (CSO) as defined by
\citet{wil94}.  The two steep-spectrum components about $0\,\farcs
002$ from the core at position angles 60\degr\ and 245\degr\ can be
interpreted as a weak jet and a counter-jet of approximately equal
flux density, so they are not relativistically beamed.  Thus the line
between them need not be close to the line of sight \citep{sch79} and
their actual lengths probably are not much greater their projected
lengths.  \citet{con92} noted that CSOs usually have compact
self-absorbed cores with spectra peaking in the GHz frequency range,
but these core sources are probably not relativistically boosted (else
triple CSOs would be outnumbered by misaligned coreless compact double
sources in radio surveys), so their flux densities can be used to
calculate their intrinsic brightness temperatures and luminosities.
The low polarization of NVSS J171719+422659 (fractional polarization
$3\sigma$ upper limit $< 0.9$\%) is also typical of CSOs \citep{gug07}.

\subsection{The Radio Galaxy Originally Identified with B3 1715+425}

\begin{figure}[htb!]
\begin{center}
\includegraphics[scale=0.45, clip=true]{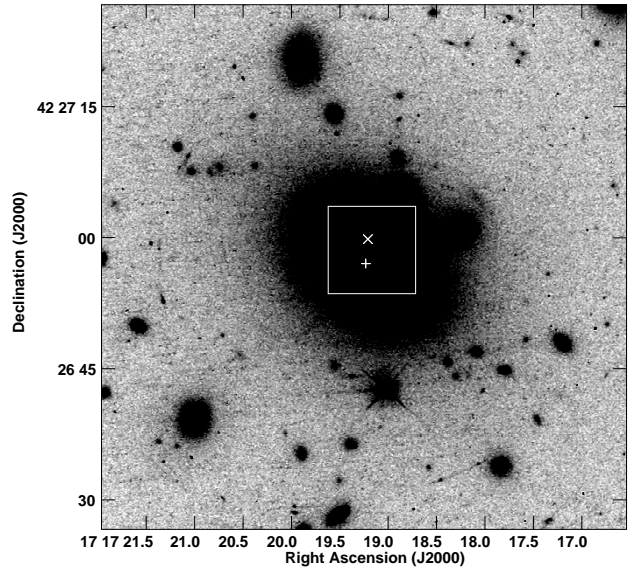}
\end{center}
\caption{This deep HST F606W image of the galaxy cluster ZwCl~8193
  shows the full extent of the BCG and some of its gravitationally
  lensed background galaxies. The white $\times$ marks the VLBA radio
  position, the white $+$ marks the HST optical position of the BCG
  core, and the white box bounds the zoomed image in Figure~\ref{fig:hstf606w}.
\label{fig:hstdeep}}
\end{figure}

The full extent of the bloated BCG in the massive cluster ZwCl~8193 is
apparent in the deep HST F606W (center $\lambda \approx 606
\mathrm{~nm}$) wideband ($480 \mathrm{~nm} < \lambda < 710
\mathrm{~nm}$) image \citep{ode10} downloaded from the Hubble Legacy
Archive (HLA) and shown in Figure~\ref{fig:hstdeep}; it is at least
$30'' \approx 90 \mathrm{~kpc}$ at $z_\mathrm{a} = 0.1754$. The $+$
marks the HST position of the BCG core and the $\times$ marks the VLBA
position of the radio source.  Several arcs of gravitationally lensed
background galaxies surround it.

\begin{figure}[htb!]
\epsscale{1.184}
\plotone{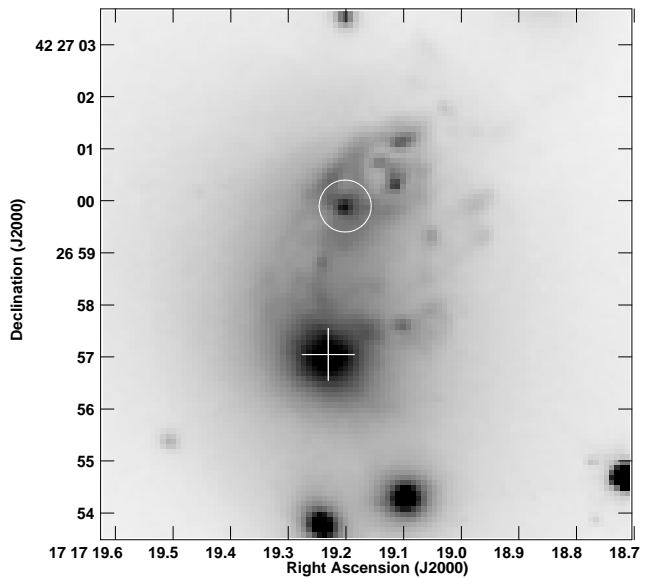}
\caption{This shallower zoomed HST F606W image of ZwCl 8183 reveals
  the core of the BCG, the radio galaxy, several cluster galaxies, and
  extended interaction debris. The HST position of the BCG core is
  marked by the white $+$, and our new optical identification of B3
  1715+425 is enclosed by a white circle centered on its HST position.
  The HST and VLBA positions coincide within $0\,\farcs1$.
 \label{fig:hstf606w}}
\end{figure}

Figure~\ref{fig:hstf606w} covers 
the white box in Figure~\ref{fig:hstdeep}
with a shallower transfer function to reveal
the core of the BCG, several cluster galaxies, and what \citet{ode10}
variously described as a dust lane plus ``apparent FUV bright debris
features,'' and thus the radio source ``may be associated with a
merging galaxy.''  On this HST image, our Gaussian fits give the BCG core
position J2000 $\alpha = 17^\mathrm{h}\,17^\mathrm{m}\, 19\,\fs 231$,
$\delta = +42^\circ\,26'\,57\,\farcs05$ and the circled object
is at J2000 $\alpha = 17^\mathrm{h}\,17^\mathrm{m}\, 19\,\fs 202$,
$\delta = +42^\circ\,26'\,59\,\farcs90$, only $\Delta \theta = 0
\,\farcs 10$ from the VLBA position. 
 
Our VLBA images reveal no radio emission brighter than 1~mJy~beam at
the position of the BCG.  FIRST \citep{bec95} ($5\,\farcs4$ FWHM
resolution) does not resolve B3 1715+425, and the FIRST flux density
$S \approx 132.5 \mathrm{~mJy}$ equals the NVSS ($45''$ FWHM
resolution) flux density $S = 133.5 \pm 4.0 \mathrm{~mJy}$, so B3
1715+425 has very little or no extended radio emission on angular
scales between $5\,\farcs4$ and $45''$.  The $2\,\farcs8$ angular
separation between B3 1715+425 and the BCG is smaller than the FIRST
beam, so the BCG might be a fairly compact radio source, smaller than
the FIRST beam but large enough to be resolved out by the VLBA.  In
that case, the FIRST position of B3 1715+425 would be at the
flux-weighted centroid of the BCG and VLBA positions.  The FIRST
declination is only $\Delta \theta = 0\,\farcs14 \pm 0\,\farcs10$
south of the VLBA declination while the BCG and VLBA positions are
separated by about $2\,\farcs8$, so the BCG likely contributes $< 5$\%
to the total flux density of B3 1715+425.

We conclude that the correct
optical identification of the radio source B3 1715+425 is \emph{only}
the circled object, and not the BCG nucleus $2\,\farcs85 \approx 8.5
\mathrm{~kpc}$ to the south or other tidal debris.  The 2MASX infrared
source is a marginally significant $\Delta \theta \sim 0\,\farcs 43$
north of the optical BCG nucleus, which is consistent with the
$\lambda = 2.16 \,\mu \mathrm{m}$ source being a blend of emission
from the BCG and material to the north.
  
The compact galaxy that we call the unique optical identification of
B3~1715+425 is listed in the HLA catalog with F606W AB magnitude $m =
21.417 \pm 0.048$ measured through a $0\,\farcs3$ radius circular
aperture.  At distance modulus $m - M = 39.64$, its absolute red
magnitude is $M_\mathrm{r} \approx -18.2$. Its concentration index is
$CI = 1.794$, so it is flagged as extended ($CI > 1.3$).  A Gaussian
fit on the HST F606W image yields a half-power ellipse $0\,\farcs39
\times 0\,\farcs30$ in size while stars of comparable magnitude appear
as $0\,\farcs25$ FWHM circles, so the deconvolved FWHM size of the
galaxy hosting B3 1715+425 is only $0\,\farcs3 \times 0\,\farcs2 \sim
0.9 \mathrm{~kpc} \, \times \, 0.6 \mathrm{~kpc}$.  Thus this radio
galaxy is much smaller and less luminous than the $(0.3 \rightarrow
1)L^*$ elliptical galaxies identified with other CSOs \citep{rea96} or
comparably luminous radio sources.

\begin{figure}[htb!]
\includegraphics[angle=270, scale=0.37,clip=true]{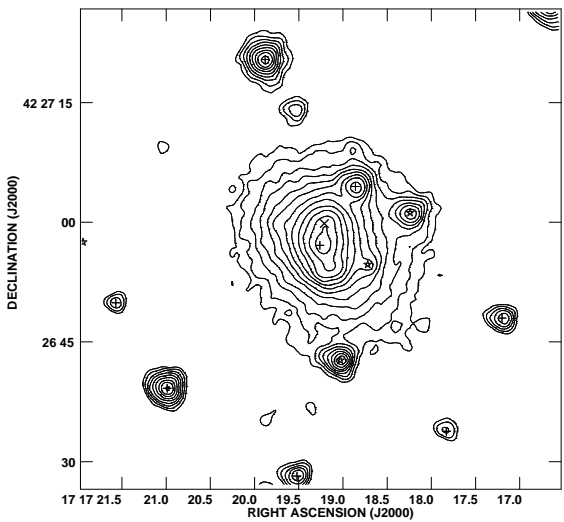}
\caption{IRAC1 ($\lambda = 3.6\,\mu\mathrm{m}$) contour image of
  ZwCl~8193.  At $\lambda = 3.6\,\mu\mathrm{m}$ the brightness is
  greatest at the position of the BCG nucleus (marked by the $+$ at
  the image center).  Successive contours are $0.10
  \mathrm{~MJy~sr}^{-1} \times 2^0,~2^{1/2},~2^1,~2^{3/2}, \dots$
  above the background.
\label{fig:irac1}}
\end{figure}

\begin{figure}[htb!]
\includegraphics[angle=270, scale=0.37,clip=true]{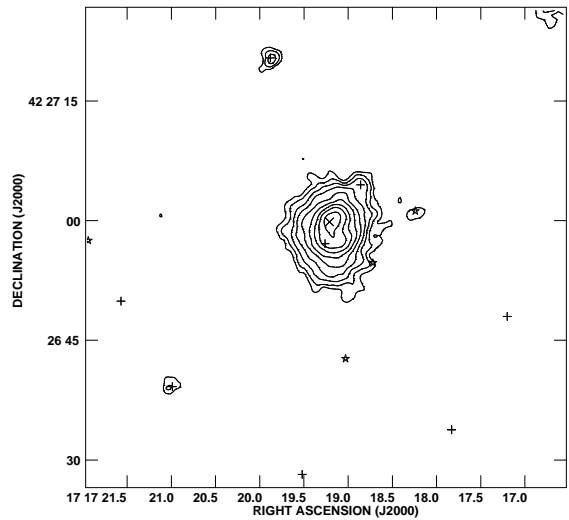}
\caption{IRAC4 ($\lambda = 8\,\mu\mathrm{m}$) contour image of Zw~8193. 
The radio source position ($\times$) is brighter than the BCG nucleus.
Successive contours are $0.25 \mathrm{~MJy~sr}^{-1}
\times 2^0,~2^{1/2},~2^1,~2^{3/2}, \dots$ above the background.
\label{fig:irac4}}
\end{figure}

\begin{figure}[htb!]
\includegraphics[angle=270, scale=0.37,clip=true]{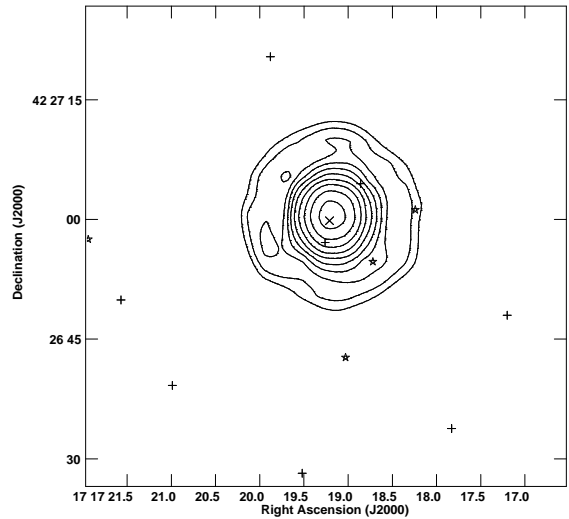}
\caption{MIPS-24 ($\lambda = 24\,\mu\mathrm{m}$, $\theta \approx 6''$
FWHM resolution) contour image of Zw~8193.  Stars and galaxies are
nearly invisible, and the centroid is near the radio position marked
by the $\times$.  Successive contours are $0.20 \mathrm{~MJy~sr}^{-1}
\times 2^0,~2^{1/2},~2^1,~2^{3/2}, \dots$ above the background.
\label{fig:mips24}}
\end{figure}

\subsection{Infrared Sources in ZwCl~8193}

The infrared structure of ZwCl~8193 is resolved in archival Spitzer
IRAC images \citep{qui08} downloaded from the NASA/IPAC Infrared
Science Archive (IRSA) in IRAC Bands 1 ($\lambda \approx
3.6\,\mu\mathrm{m}$, $\theta \approx 1\,\farcs9$ FWHM resolution) and
4 ($\lambda \approx 8\,\mu\mathrm{m}$, $\theta \approx 2\,\farcs8$
FHWM), and it is marginally resolved in the MIPS-24 ($\lambda \approx
24\,\mu \mathrm{m}$, $\theta \approx 6\,\farcs4$ FWHM) image.
Infrared contour plots covering the same area as
Figure~\ref{fig:hstdeep} are shown in Figures~\ref{fig:irac1},
\ref{fig:irac4}, and \ref{fig:mips24}.  The near- and mid-infrared
images help to distinguish between Galactic stars ($\star$), cluster
galaxies ($+$), and the radio galaxy ($\times$).  Individual stars and
dustless galaxies of stars have temperatures $T \gtrsim 3500
\mathrm{~K}$ and Wien peak wavelengths $\lambda_\mathrm{W} \lesssim
1\,\mu\mathrm{m}$, so they are significantly brighter in Band 1 than
in Band 4, and they have disappeared in MIPS-24.  In Band 1, the BCG
nucleus ($+$) is brighter than the radio galaxy ($\times$), but the
radio source position marks the highest brightness in the whole Band 4
field, and at $\lambda = 24 \,\mu \mathrm{m}$ it is completely
dominant, as indicated by the MIPS-24 position J2000 $\alpha =
17^\mathrm{h}\,17^\mathrm{m}\, 19\,\fs17$, $\delta= +42^\circ 27'
00\,\farcs5$ determined by a Gaussian fit.  \citet{qui08} noted that
ZwCl~8193 has a strong mid-IR excess compared with a quiescent
elliptical galaxy.  Such excess emission is usually produced by warm
interstellar dust at temperatures $T \ll 3500 \mathrm{~K}$.  We
identify the mid-infrared source in ZwCl~8193 \emph{only} with the
radio galaxy and its surrounding debris, and not with the elliptical
BCG in ZwCl~8193.

From the MIPS image we measured a $\lambda = 24\,\mu \mathrm{m}$ ($\nu
= 1.25 \times 10^{13} \mathrm{~Hz}$) flux density
$S_{24\,\mu\mathrm{m}} \approx 10 \pm 1 \mathrm{~mJy}$.  The
$\lambda = 70 \, \mu \mathrm{m}$ ($\nu \approx 4.3 \times 10^{12}
\mathrm{~Hz}$) flux density of the radio galaxy is
$S_{70\,\mu\mathrm{m}} \approx 177 \mathrm{~mJy}$ \citep{qui08}, and
the ratio $S_{70\,\mu\mathrm{m}} / S_{24\,\mu\mathrm{m}} \sim 20$ is
typical of galaxies selected at $\lambda = 70\,\mu\mathrm{m}$
\citep{cle11}.  Thus the mid-infared luminosity of the radio galaxy is
$\nu L_\nu \approx 3 \times 10^{10} L_\odot$ at $\lambda = 24\,\mu
\mathrm{m}$, and its total infrared luminosity may be up to $10
\times$ higher, $L_\mathrm{IR} \sim 3 \times 10^{11} L_\odot$. We
cannot rule out a minor contribution by the BCG to the $70\,\mu
\mathrm{m}$ luminosity because the $70\,\mu \mathrm{m}$ resolution is
inadequate ($\theta \approx 20''$ FWHM) to distinguish the radio
galaxy from the BCG.

The steep spectrum between $24 \,\mu \mathrm{m}$ and $70 \,\mu
\mathrm{m}$ indicates a low dust temperature $T_\mathrm{d} < 100
\mathrm{~K}$.  The specific intensity of the dust source cannot exceed
that of a $T = T_\mathrm{d} < 100 \mathrm{~K}$ blackbody, so the
combination of high luminosity and low temperature implies that the
diameter of the dust source is at least
\begin{equation}
D_\mathrm{d} >
\Biggl(\frac {L_\mathrm{IR}} { \pi \sigma T_\mathrm{d}} \Biggr)^{1/2}
\sim 100 \mathrm{~pc}~,
\end{equation}
where $\sigma \approx 5.67 \times 10^{-5} \mathrm{~erg~s}^{-1}
\mathrm{~cm}^{-2} \mathrm{~K}^{-4}$ is the Stefan-Boltzmann constant.
Thus the dust source is much larger than the SMBH accretion disk. The dust
may well be heated by the SMBH; the stars in the optically faint radio
galaxy and surrounding debris can contribute significantly to the high
infrared luminosity only if they are young, massive, and heavily
obscured at visible wavelengths, as in the case of Arp 220.

\section{X-ray Sources}
\label{sec:xray}

The X-ray source 1RXS J171718.9+422652 at J2000 $\alpha =
17^\mathrm{h} \, 17^\mathrm{m} \, 18\,\fs 9$, $\delta = +42^\circ \,
26' \, 52''$ overlaps ZwCL~8193 \citep{mas09}, the BCG, and the radio
galaxy. Its 0.1--2.4 keV flux $2.03 \times 10^{-12}
\mathrm{~erg~cm}^{-2}$ corresponds to a soft X-ray luminosity $\approx
1.7 \times 10^{44} \mathrm{~erg~s}^{-1}$ at the 850~Mpc luminosity
distance of the cluster.

We observed ZwCl 8193 with the {\it Chandra X-ray Observatory} on 2013
October 7 using the ACIS S3 chip in a timed exposure ``faint'' mode
as part of program GO3-14112X.  The nominal integration time was 18202
s.  We reprocessed and reduced the observations using CIAO 4.8 and
calibration files CALDB 4.7.2.  Data were filtered for background
flares and restricted to the energy range 0.5--7 keV.

\begin{figure}[htb!]
\includegraphics[angle=0, scale=0.46,clip=true]{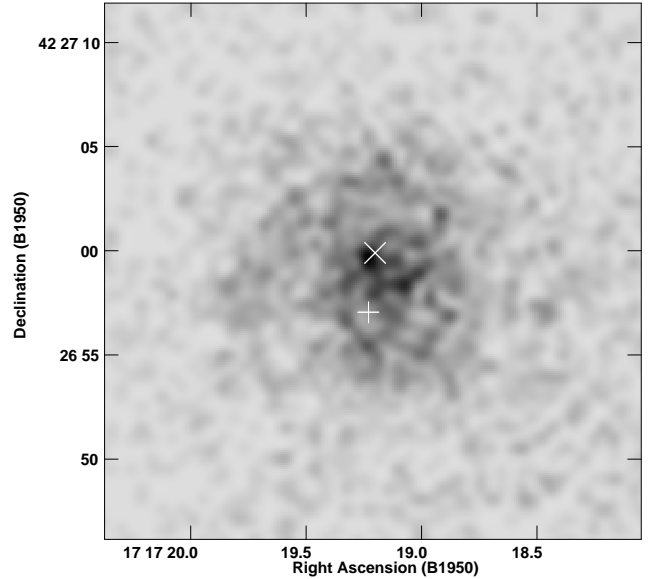}
\caption{
The white $\times$ marks the VLBA radio position of B3 1715+425, and the
white $+$ marks the HST optical position of the BCG core in this 
{\it Chandra} 0.5--7 keV X-ray image.  The detection of the unresolved
X-ray source on the radio position is significant, and there is no compact
source visible on the BCG position.
\label{fig:chandra}}
\end{figure}

Figure~\ref{fig:chandra} shows the 0.5--7 keV {\it Chandra} image with
the radio source and BCG marked.  An X-ray peak coincides with the
radio source and is clearly distinguishable from the extended X-ray
emission associated with the cluster or BCG.  Wavelet detection
succeeds in identifying the unresolved X-ray source with centroid
J2000 $\alpha = 17^\mathrm{h}\, 17^\mathrm{m}\, 19\,\fs 22$, $\delta =
+42^\circ \,26' \, 59\,\farcs6$, but measuring the X-ray counts in the
presence of the highly structured extended emission required custom
photometry that is highly uncertain.  We therefore did not attempt to
model the source spectrum or correct for absorption.

While the X-ray properties of the BCG are highly uncertain, the
detection of an unresolved X-ray source coincident with the radio
source is significant.  We detected $85\pm15$ counts and a count rate
of $0.0047\pm 0.0008 \mathrm{~s}^{-1}$ in the 0.5--7 keV range
(statistical errors only).  The flux is $(3.6\pm 0.6)\times 10^{-14}
\mathrm{~erg~cm}^{-2} \mathrm{~s}^{-1}$, and the 0.5--7 keV X-ray
luminosity is $(3.1\pm0.5)\times 10^{42} \mathrm{~erg~s}^{-1}$ (again,
all errors are statistical and do not include the systematic errors
arising from separating the compact from the extended emission).

\section{Long-slit Spectroscopy of ZwCl~8193}
\label{sec:spectra}

\begin{figure}[htb!]
\epsscale{1.}
\plotone{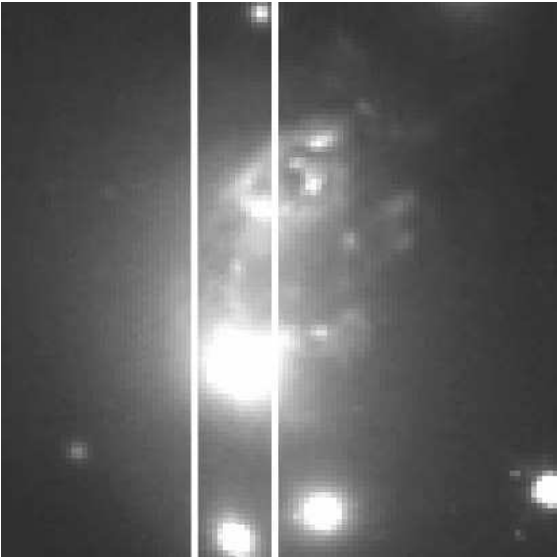}
\caption{The $1\,\farcs5$ wide spectrograph slit between the vertical
  white lines includes the radio galaxy, the BCG, and the region
  between them. \label{fig:slit}}
\end{figure}

After \citet{all92} reported an emission-line redshift $z_\mathrm{e} =
0.1829$ that they associated with the BCG in ZwCl~8193, \citet{edg02}
imaged the faint Pa$\alpha$ emission connecting the BCG and B3
1715+425 and estimated a velocity difference between them of $\sim
400-500 \mathrm{~km~s}^{-1}$.  We observed ZwCl~8193 using the Double
Imaging Spectrograph on the Astrophysical Research Consortium (ARC)
3.5 m telescope at the Apache Point Observatory on 17 April 2012.  The
long slit was $1\,\farcs5$ wide and oriented north-south to include
both B3 1715+425 and the BCG (Figure~\ref{fig:slit}).  We obtained red
(resolution $R\sim3250$) and blue ($R\sim2400$) spectra from the UV
atmospheric cutoff to $9800\, \mathrm{\AA}$ that included
H$\alpha$6563, the [N\,\textsc{ii}]6549,6583 doublet, H$\beta$4861,
the [O\,\textsc{iii}]4959,5007 doublet, and [O\,\textsc{iii}]4363 in
emission plus the Ca\,\textsc{ii} H and K $\lambda =3968,~3934\,
\mathrm{\AA}$ lines in absorption.  Eight 600 s exposures were
flat-fielded, wavelength-calibrated, rectified, sky-subtracted, and
median combined.  The absolute velocity uncertainties are $\sim150$
km~s$^{-1}$, but the relative uncertainties are only $\sim10$
km~s$^{-1}$ for bright emission lines.  The Ca\,\textsc{ii} H and K
lines confirm the Allen et al.\ (1992) absorption line redshift toward
the BCG, and the [O\,\textsc{iii}]4363 line confirms the reported
offset between absorption-line and emission-line redshifts.

\begin{figure}[htb!]
\epsscale{1.}
\plotone{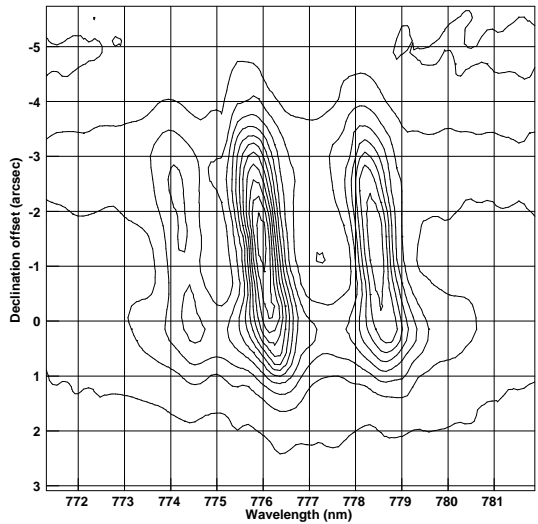}
\caption{Contour plot of the emission-line spectrum centered on the
  H$\alpha$ and adjacent [N\,\textsc{ii}] lines.  Successive contours
  are separated by the same linear interval in brightness (arbitrary
  units).  Abscissa: Wavelength (nm).  Ordinate: Declination offset
  from the optical counterpart of B3 1717+425; note that north is
  down.  The BCG nucleus is at $\Delta \delta = -2\,\farcs85$, and its
  continuum emission is visible as the horizontal band centered on
  that declination offset. \label{fig:redspectrum}}
\end{figure}

A contour plot of the H$\alpha$ and [N\,\textsc{ii}] line intensities
is shown in Figure~\ref{fig:redspectrum}, with north pointing down.
The H$\alpha$ emission-line redshift varies from $z_1 \approx 0.18215$
near the BCG nucleus at $\Delta \delta \approx -2\,\farcs85$ to $z_2
\approx 0.18271$ at the declination of B3 1715+425 ($\Delta \delta
\equiv 0$), for a redshift difference $\Delta z = (z_2 - z_1) / (1 +
z_1) \approx 0.00047$ and a radial velocity difference of only $v
\approx cz \approx 140 \mathrm{~km~s}^{-1}$.  The difference between
the emission-line redshift $z_\mathrm{e} = 0.18271$ of B3 1715+425 and
the absorption-line redshift $z_\mathrm{a} = 0.1754$ of the BCG galaxy
is $\Delta z_\mathrm{ea} = (z_\mathrm{e} - z_\mathrm{a}) / (1 +
z_\mathrm{a}) \approx 0.0062$, so the ionized gas in B3 1715+425 is
moving away from the BCG with a radial speed $v_\mathrm{r} \approx cz
\approx 1860 \mathrm{~km~s}^{-1} \gg 140 \mathrm{~km~s}^{-1}$.  The
FWHM width of the H$\alpha$ line is $\Delta v \approx 480
\mathrm{~km~s}^{-1}$ near B3 1715+425 and falls to $\Delta v \lesssim
320 \mathrm{~km~s}^{-1}$ midway between B3 1715+425 and the BCG.
These features are very similar to those of the disk and ionized tail
of the galaxy ESO~137-001 that is undergoing ram-pressure stripping as
it falls into the Norma cluster \citep{fos16}, so all of the
emission-line gas appears to have originated in B3 1715+425, and we
use $v_\mathrm{r} \approx 1860 \mathrm{~km~s}^{-1}$ as the radial
velocity of B3 1715+425 relative to the BCG.

The diagnostic emission-line intensity ratios $\log(
\mathrm{[N\,\textsc{ii}]/H}\alpha) \approx -0.2$ and $\log
(\mathrm{[O\,\textsc{iii}]/H}\beta) \approx -0.15$ lie in the
``composite'' range of the BPT diagram \citep{kew06}, suggesting that
much of the line emission is from photoionization by stars. The far
weaker [O\,\textsc{iii}]4363 emission indicates little ionization by
strong shocks.

\section{Interpretation}
\label{sec:interpretation}

We unambiguously identify the compact and luminous radio source 
B3 1715+425 with a uniquely small and optically faint radio galaxy at
least 8.5~kpc from the nucleus of the BCG in ZwCl~8193.  An ionized
wake extends from the radio galaxy to the BCG, suggesting that the
radio galaxy is now moving away from the BCG with a radial component
of velocity $v_\mathrm{r} \approx 1860 \mathrm{~km~s}^{-1}$.  BCGs can
have nuclear escape velocities up to $3000 \mathrm{~km~s}^{-1}$
\citep{mer04}, so the fleeing radio galaxy may still be
gravitationally bound to the BCG and/or the cluster ZwCl~8193.

The SMBH powering the radio source is still $\lesssim 0.1
\mathrm{~kpc}$ from the radio galaxy nucleus, so it must have nearly
the same velocity as its host emission-line galaxy.  The high SMBH
velocity relative to the BCG is consistent with (1) gravitational
recoil following the merger of two spinning SMBHs or slingshot
ejection of one SMBH from a triple system in the BCG or (2) the
velocity of a galaxy that fell into the gravitational potential well
in the center of the massive cluster ZwCl~8193.

In case (1), tidal forces would have stripped the SMBH of its
surrounding stars and dark matter just prior to a merger
\citep{hof07}.  The violent jerk of post-merger gravitational-wave recoil or
slingshot ejection would have left behind most of what remained, so
the initially naked SMBH would have to accrete the small but finite
($\sim 0.9 \mathrm{~kpc}\times 0.6 \mathrm{~kpc}$) radio galaxy we see
today while leaving the BCG at speeds $v \gtrsim 2 \times 10^{\,3}
\mathrm{~km~s}^{-1}$.  The speeding SMBH can accrete only the stars,
dark matter, and gas that lie within an impact parameter $b$ below
which the gravitational potential energy $(G M / b)$ of the SMBH is
greater than the kinetic energy per unit mass $v^2/ 2$ of the accreted
matter, so
\begin{equation} 
\label{eqn:accretion}
\biggl( \frac {b} {\mathrm{pc}}\biggr) \approx
2.2  \biggl( \frac {M} {10^{\,9} M_\odot} \biggr) 
\biggl( \frac {v}{2000 \mathrm{~km~s}^{-1}} \biggr)^{-2}~.
\end{equation}
The velocity dispersion or rotational velocity of the accreted
material is only $v_\mathrm{rot} \approx \Delta v / 2 \approx 240
\mathrm{~km~s}^{-1}$ at radial distance $r \approx 450 \mathrm{~pc}$
from the SMBH, indicating that the total mass of the radio galaxy
including its SMBH is not more than $M \approx r v_\mathrm{rot}^2 /
G$, or
\begin{equation}
\label{eqn:rotation}
\biggl( \frac {M}{10^{\,9} M_\odot}\biggr) \lesssim 6.0
\biggl( \frac{r}{450 \mathrm{~pc}} \biggr)
\biggl( \frac{v_\mathrm{rot}}{240 \mathrm{~km~s}^{-1}} \biggr)^2~.
\end{equation}
Inserting $M \lesssim 6 \times 10^{\,9} M_\odot$ from
Equation~\ref{eqn:rotation} into Equation~\ref{eqn:accretion} yields
$b \lesssim 10 \mathrm{~pc}$, which is much smaller than the radius of
the galaxy surrounding the SMBH.  
% The entence below has been deleted: 
% Also, the recoiling SMBH should have
% decelerated as it left the BCG nucleus, so the H$\alpha$ velocity
%g radient should have been opposite to what is observed.  
We conclude
that any SMBH violently ejected from the BCG nucleus with velocity $v
\gtrsim 2000 \mathrm{~km~s}^{-1}$ could not have accreted the extended
galaxy that currently surrounds it.

This leaves case (2), a formerly normal galaxy that fell into
ZwCl~8193, was tidally stripped as it passed near the BCG nucleus, and
is now rapidly moving away from the BCG.  Only the SMBH and the small
central portion of that galaxy denser than the BCG core remain intact,
with total mass $M \lesssim 6 \times 10^{\,9} M_\odot$ and radius $r \sim
450 \mathrm{~pc}$.  Such severe stripping requires a relatively 
rare 
collision with impact parameter $b \lesssim 1\mathrm{~kpc}$, the
core radius of the most massive elliptical galaxies \citep{fab97}.
The path of the escaping galaxy is traced by its
ionized tail, the velocity of the escaping galaxy is probably close to
the cluster escape velocity, and the small velocity gradient along
that tail is caused by gas drag from the ISM of the BCG.  The outer
portion of the stripped galaxy now appears as ``debris'' north of the
BCG nucleus in Figure~\ref{fig:hstf606w}.

Finally, most massive bulges contain nuclear SMBHs, and the typical
bulge/SMBH mass ratio is $\sim 500$ \citep{mcc13}.  Heavy stripping
significantly reduces this ratio, making low-mass stripped galaxies
surrounding massive SMBHs more vulnerable to additional mass loss
caused by ``quasar mode'' radiative feedback.  The $L_\mathrm{IR} \sim
3 \times 10^{11} L_\odot$ total infrared luminosity of our $M \lesssim
6 \times 10^{\,9} M_\odot$ stripped galaxy is well below its classiscal
Eddington limit $(L_\mathrm{E} / L_\odot) \approx 3.3 \times 10^{\,4} (M /
M_\odot) \lesssim 2 \times 10^{14}$ caused by radiation pressure from
Thomson scattering in ionized hydrogen gas.  However, the radiation
pressure per unit mass from dust scattering can be up to 500 times
larger \citep{fab08}, so the luminosity of our stripped galaxy may be
close to its dust Eddington limit $L_\mathrm{D} \lesssim 4 \times
10^{11} L_\odot$.

\acknowledgments This work is based on observations made at the Apache
Point Observatory, operated by New Mexico State University.  The
  scientific results reported in this article are based on
  observations made with the Chandra X-ray Observatory.  Support for
  this work was provided by the National Aeronautics and Space
  Administration through Chandra Award Number GO3-14112X issued by the
  Chandra X-ray Observatory Center, which is operated by the
  Smithsonian Astrophysical Observatory for and on behalf of the
  National Aeronautics Space Administration under contract NAS8-03060.
  This research has made use of software provided by the Chandra X-ray
  Center (CXC) in the application packages CIAO and ChIPS.  This
research has made use of the NASA/IPAC Extragalactic Database (NED)
and the NASA/\allowbreak IPAC Infrared Science Archive (IRSA), which
are operated by the Jet Propulsion Laboratory, California Institute of
Technology, under contract with the National Aeronautics and Space
Administration.  The Hubble Legacy Archive (HLA) is based on
observations made with the NASA/ESA Hubble Space Telescope, and
obtained from the Hubble Legacy Archive, which is a collaboration
between the Space Telescope Science Institute (STScI/NASA), the Space
Telescope European Coordinating Facility (ST-ECF/ESA) and the Canadian
Astronomy Data Centre (CADC/\allowbreak NRC/\allowbreak CSA).

{\it Facilities:} \facility{ARC}, \facility{CXO}, \facility{VLBA}.

Data for this paper are provided in a persistent repository
\dataset [ADS/Sa.CXO#obs/14988] {Chandra ObsId 14988}

\clearpage

\end{document}